\documentclass{article}
\usepackage[a4paper, portrait, margin=1.1811in]{geometry}
\usepackage[english]{babel}
\usepackage[utf8]{inputenc}
\usepackage[T1]{fontenc}
\usepackage{helvet}
\usepackage{etoolbox}
\usepackage{graphicx}
\usepackage{titlesec}
\usepackage{caption}
\usepackage{booktabs}
\usepackage{xcolor}
\usepackage{multirow}
\usepackage{multicol}
\usepackage{natbib}
\usepackage[colorlinks, citecolor=cyan]{hyperref}
\usepackage{amsmath, amsfonts, amssymb, amsthm, graphicx, enumerate, enumitem, multicol}
\usepackage{caption}
\graphicspath{ {./images/} }
\usepackage{scrextend}
\usepackage{fancyhdr}

\fancypagestyle{plain}{
	\fancyhf{}

}

\makeatletter
\patchcmd{\@maketitle}{\LARGE \@title}{\fontsize{16}{19.2}\selectfont\@title}{}{}
\makeatother

\usepackage{authblk}

\newsavebox\affbox
\author[1]{\textbf{Benjamin Osafo Agyare}}
\affil[1]{Department of Statistics, University of Michigan, Ann Arbor.
}
\titlespacing\section{0pt}{12pt plus 4pt minus 2pt}{0pt plus 2pt minus 2pt}
\titlespacing\subsection{12pt}{12pt plus 4pt minus 2pt}{0pt plus 2pt minus 2pt}
\titlespacing\subsubsection{12pt}{12pt plus 4pt minus 2pt}{0pt plus 2pt minus 2pt}

\titleformat{\section}{\normalfont\fontsize{15}{15}\bfseries}{\thesection.}{1em}{}
\titleformat{\subsection}{\normalfont\fontsize{12}{15}\bfseries}{\thesubsection.}{1em}{}
\titleformat{\subsubsection}{\normalfont\fontsize{10}{15}\bfseries}{\thesubsubsection.}{1em}{}

\titleformat{\author}{\normalfont\fontsize{12}{15}\bfseries}{\thesection}{1em}{}

\title{\textbf{Enhancing Computational Efficiency in High-Dimensional Bayesian Analysis: Applications to Cancer Genomics}
}
\date{}    

\begin{document}

\pagestyle{headings}	
\newpage
\setcounter{page}{1}
\renewcommand{\thepage}{\arabic{page}}

\captionsetup[figure]{labelfont={bf},labelformat={default},labelsep=period,name={Figure }}	\captionsetup[table]{labelfont={bf},labelformat={default},labelsep=period,name={Table }}
\setlength{\parskip}{0.5em}
	
\maketitle
	
\noindent\rule{15cm}{0.5pt}
	\begin{abstract}
		%High-dimensional data poses significant challenges in linear regression, particularly when the number of covariates ($p$) far exceeds the sample size ($n$). A common frequentist solution is to use penalization techniques for shrinkage; however, these methods often complicate the quantification of parameter uncertainty. The Bayesian framework offers an alternative by facilitating uncertainty quantification through the estimation of posterior distributions, typically via Markov Chain Monte Carlo (MCMC) techniques. In high-dimensional settings, efficient and computationally feasible MCMC algorithms are crucial. This simulation study compares the performance of the two-Block Gibbs sampler (2BG) with the three-Block Gibbs sampler (3BG) in estimating the posterior distributions for two widely-used Bayesian shrinkage models: the Bayesian Lasso (BL) and Spike-and-Slab priors. The comparison is based on metrics such as one-lag autocorrelation and the average effective sample size per second, $N_{\mathrm{eff}}/T$. We demonstrate that the 2BG is more efficient and computationally advantageous compared to the 3BG. Finally, we apply these methods to protein expression genetics data from the National Cancer Institute to illustrate their practical utility.\\

        In this study, we present a comprehensive evaluation of the Two-Block Gibbs (2BG) sampler as a robust alternative to the traditional Three-Block Gibbs (3BG) sampler in Bayesian shrinkage models. Through extensive simulation studies, we demonstrate that the 2BG sampler exhibits superior computational efficiency and faster convergence rates, particularly in high-dimensional settings where the ratio of predictors to samples is large. We apply these findings to real-world data from the NCI-60 cancer cell panel, leveraging gene expression data to predict protein expression levels. Our analysis incorporates feature selection, identifying key genes that influence protein expression while shedding light on the underlying genetic mechanisms in cancer cells. The results indicate that the 2BG sampler not only produces more effective samples than the 3BG counterpart but also significantly reduces computational costs, thereby enhancing the applicability of Bayesian methods in high-dimensional data analysis. This contribution extends the understanding of shrinkage techniques in statistical modeling and offers valuable insights for cancer genomics research.\\

		\textbf{\textit{Keywords}}: \textit{Bayesian Lasso; Spike-and-Slab; Gibbs sampler; High-Dimensional Statistics, MCMC Methods, Cancer Genomics}
	\end{abstract}
\noindent\rule{15cm}{0.4pt}

\section{Introduction}
Linear regression aims to model the relationship between a response variable, \(y\), and a set of predictor variables, \(x_1, x_2, \dots, x_p\), using regression coefficients \(\boldsymbol{\beta} = (\beta_1, \beta_2, \dots, \beta_p)\). In situations where the number of predictors, \(p\), far exceeds the number of observations, \(n\), penalization techniques are often employed. These methods help to regularize the model by applying a penalty to the regression coefficients, effectively shrinking some \(\beta_j\)’s towards zero. This leads to sparse estimates, which can improve model stability and prediction accuracy in high-dimensional settings. However, despite their effectiveness, penalization methods face the challenge of providing uncertainty quantification for the estimated parameters.

In the Bayesian paradigm, inference is achieved by incorporating prior distributions on the parameters, allowing for direct quantification of uncertainty through the posterior distribution. In recent years, Bayesian shrinkage models have gained popularity for addressing high-dimensional problems, with two main approaches: the use of \textit{spike-and-slab priors}, and continuous shrinkage priors. Spike-and-slab models, such as Laplace-Zero mixtures \citep{johnstone2004needles}, combine a point mass at zero with a continuous distribution, providing a flexible framework for inducing sparsity. On the other hand, \textit{purely continuous shrinkage priors}, such as the Horseshoe prior \citep{carvalho2010horseshoe}, have emerged as effective alternatives for handling high-dimensional data while quantifying parameter uncertainties.

Additionally, the Bayesian lasso, introduced by \cite{park2008bayesian}, extends the frequentist lasso by interpreting its objective as a Bayesian model. This approach places an independent Laplace prior on the regression coefficients, connecting it to the original lasso formulation developed by \cite{tibshirani1996regression}, which has been widely used for regularization and variable selection in high-dimensional regression.

However, as with most Bayesian problems, there are no closed form expression for posterior quantification of the Bayesian Shrinkage models. An approach to approximating the posterior distribution is via Variational Bayes algorithms such as mean field variational Bayes (MFVB) methodology. While this is a fast algorithm, \cite{neville2014mean} points that the MFVB algorithm can perform quite poorly due to posterior dependence among auxiliary variables. Markov chain Monte Carlo (MCMC), an indispensable tool in contemporary Bayesian inference provides us with alternative methods that can be used to approximate the posterior distribution and hence make inferences on the parameters, since the shrinkage priors of these models capitalize on hierarchical representations. Utilizing the power of MCMC, \cite{park2008bayesian} provided the three-step Gibbs sampler, otherwise known as the three-Block Gibbs sampler (3BG) algorithm to explore the posterior distribution for arbitrary $n$ and $p$. While \cite{khare2013geometric} proved that the 3BG is geometrically ergodic for arbitrary values of $n$ and $p$, it tends to suffer from slow convergence especially if $p/n$ is large enough due to high correlation between components of the different blocks, especially that between the regression coefficients in one block and the variance parameters in another\citep{rajaratnam2015mcmc}. Owing to the deterioration in the convergence properties of the 3BG in high-dimesional settings, \cite{Rajaratnam2019} proposed an efficient version known as the two-Block Gibbs sampler (2BG) algorithm to overcome the sluggish convergence issues of the former. Theoretical convergence properties can be found in the aforementioned paper for interested readers.

In this study, an extensive simulation analysis is conducted to compare the computational efficiencies of the 3BG and the 2BG using both the \textit{spike-and-slab} priors and Bayesian lasso model following \cite{Rajaratnam2019}. Here, computing efficiency is measured as the effective sample size per second ($N_{\mathrm{eff}}/T$). Further, the mixing rates are evaluated by examining lag-one autocorrelations $\rho_1$, with values closer to zero indicating improved mixing performance \citep{rajaratnam2015mcmc}. We consequently apply this method on the well-known \href{https://discover.nci.nih.gov/cellminer/}{NCI-60 cancer cell panel} from the National cancer Institute.

\cite{jin2021fast} introduced the 2BG for Bayesian group lasso, sparse group lasso, and fused lasso models, comparing its performance to the 3BG and Hamiltonian Monte Carlo (HMC). While their work focused on group-structured priors, this study takes a different direction by evaluating the performance of the 2BG relative to the 3BG for two classic Bayesian shrinkage models: the spike-and-slab prior and the Bayesian lasso. Furthermore, this paper extends the application to real-world data from the NCI-60 cancer cell line panel, using gene expression data to predict protein expression levels. By incorporating feature selection, the analysis not only identifies the most influential genes but also provides insights into the genetic mechanisms driving protein expression in cancer cells. This contribution deepens the understanding of shrinkage methods in high-dimensional settings while offering novel applications in cancer genomics.

The rest of the article is organized as follows: Section \ref{sec:methods} presents an overview of the Bayesian shrinkage framework for regression, including a review of the spike-and-slab priors, the Bayesian lasso, and the 2BG algorithms for exploring posterior distributions. Sections \ref{sec:simulation} and \ref{sec:application} detail the simulation studies and real data analyses conducted to empirically compare the two algorithms. The article concludes with a discussion of the findings in Section \ref{sec:discussion}. The codes for this project are found at https://github.com/bosafoagyare/2-BGS.

\section{Methods}\label{sec:methods}

\subsection{Bayesian Shrinkage Models}
Consider a dataset \(\mathcal{D}_n = \{(\boldsymbol{x}_1, y_1), \dots, (\boldsymbol{x}_n, y_n)\} = (\mathcal{X}_n, \mathcal{Y}_n)\), where \(\mathcal{Y}_n \in \mathbb{R}^n\) represents the response vector, and \(\mathcal{X}_n \in \mathbb{R}^{n \times p}\) is an \(n \times p\) design matrix of standardized covariates. Let \(\boldsymbol{\beta} \in \mathbb{R}^p\) denote the vector of regression coefficients, \(\sigma^2 > 0\) the residual variance, and \(\mu \in \mathbb{R}\) an unknown intercept. The regression model is expressed as:

\begin{equation} \label{eqn:1}
\boldsymbol{Y} \mid \boldsymbol{\beta}, \sigma^2 \sim \mathrm{N}_n\left(\mu \mathbf{1}_n + \boldsymbol{X} \boldsymbol{\beta}, \sigma^2 \boldsymbol{I}_n\right)
\end{equation}

The focus here is on high-dimensional regression problems where the number of covariates, \(p\), greatly exceeds the sample size, \(n\). In a Bayesian framework where sparsity is desired, the use of \textit{spike-and-slab} priors—combining a spike at zero with a normal density and another normal density that is flat near zero—has been explored as a method to shrink regression coefficients toward zero. Alternatively, \textit{purely continuous} shrinkage priors have also been used to achieve this effect. In many Bayesian high-dimensional regression settings, the prior is typically specified as:

\begin{equation}\label{eqn:2}
\boldsymbol{\beta} \mid \sigma^2, \boldsymbol{\tau} \sim \mathrm{N}_p\left(\mathbf{0}_p, \sigma^2 \boldsymbol{D}_{\boldsymbol{\tau}}\right), \quad \boldsymbol{\tau} \sim \pi(\boldsymbol{\tau}),
\end{equation}

where \(\pi(\boldsymbol{\tau})\) represents the prior distribution on \(\tau = \left(\tau_1, \ldots, \tau_p\right)\). Additionally, the prior for \(\sigma^2\) and \(\mu\) is assumed to be the improper prior \(\pi(\sigma^2, \mu) = 1 / \sigma^2\), which is independent of \(\pi(\boldsymbol{\tau})\). By integrating out \(\mu\) and combining Equations \ref{eqn:1} and \ref{eqn:2}, the following conditional distributions are obtained:

\begin{equation}\label{eqn:3}
\begin{aligned}
& \boldsymbol{\tau} \mid \boldsymbol{\beta}, \sigma^2, \boldsymbol{Y} \sim \pi\left(\boldsymbol{\tau} \mid \boldsymbol{\beta}, \sigma^2, \boldsymbol{Y}\right), \\
&  \sigma^2 \mid \boldsymbol{\beta}, \boldsymbol{\tau}, \boldsymbol{Y} \sim \text { Inverse-Gamma } \\
& {\left[(n+p-1) / 2,\|\tilde{\boldsymbol{Y}}-\boldsymbol{X} \boldsymbol{\beta}\|_2^2 / 2+\boldsymbol{\beta}^{\mathrm{T}} \boldsymbol{D}_{\boldsymbol{\tau}}^{-1} \boldsymbol{\beta} / 2\right], } \\
& \boldsymbol{\beta} \mid \sigma^2, \boldsymbol{\tau}, \boldsymbol{Y} \sim \mathrm{N}_p\left(\boldsymbol{A}_{\boldsymbol{\tau}}^{-1} \boldsymbol{X}^{\mathrm{T}} \tilde{\boldsymbol{Y}}, \sigma^2 \boldsymbol{A}_{\boldsymbol{\tau}}^{-1}\right),
\end{aligned}
\end{equation}

where \(\boldsymbol{A}_{\boldsymbol{\tau}} = \boldsymbol{X}^{\mathrm{T}} \boldsymbol{X} + \boldsymbol{D}_{\boldsymbol{\tau}}^{-1}\) and \(\boldsymbol{D}_{\boldsymbol{\tau}} = \operatorname{Diag}\left(\tau_1, \tau_2, \ldots, \tau_p\right)\).

Given the ability to sample from \(\pi(\boldsymbol{\tau} \mid \boldsymbol{\beta}, \sigma^2, \boldsymbol{Y})\), these conditional distributions can be used to construct a three-Block Gibbs sampler (3BG) for drawing from the joint posterior \(\pi(\boldsymbol{\beta}, \sigma^2 \mid \boldsymbol{Y})\). The one-step transition density \(\hat{k}\) with respect to Lebesgue measure on \(\mathbb{R}^p \times \mathbb{R}_+\) is given by:

\begin{equation}\label{eqn:4}
\begin{aligned}
\hat{k}\left[\left(\boldsymbol{\beta}_0, \sigma_0^2\right), \left(\boldsymbol{\beta}_1, \sigma_1^2\right)\right] = & \int_{\mathbb{R}_+^p} \pi\left(\sigma_1^2 \mid \boldsymbol{\beta}_1, \boldsymbol{\tau}, \boldsymbol{Y}\right) \pi\left(\boldsymbol{\beta}_1 \mid \boldsymbol{\tau}, \sigma_0^2, \boldsymbol{Y}\right) \\
& \times \pi\left(\boldsymbol{\tau} \mid \boldsymbol{\beta}_0, \sigma_0^2, \boldsymbol{Y}\right) d\boldsymbol{\tau}.
\end{aligned}
\end{equation}

\subsubsection{\textit{The Spike-and-Slab} Prior}
Under the \textit{spike-and-slab} prior framework, each \(\tau_j\) is assigned an independent discrete prior that places probability \(w_j\) on the value \(\kappa_j \zeta_j\) and probability \(1 - w_j\) on the value \(\zeta_j\), where \(\zeta_j > 0\) is a small value, \(\kappa_j > 0\) is large, and \(w_j \in (0,1)\). This setup leads to a conditional posterior distribution for \(\boldsymbol{\tau} \mid (\boldsymbol{\beta}, \sigma^2, \boldsymbol{Y})\), which is a product of independent discrete distributions. Each distribution assigns probability \(\tilde{w}_j\) to the point \(\kappa_j \zeta_j\) and probability \(1 - \tilde{w}_j\) to the point \(\zeta_j\) \citep{Rajaratnam2019}. The value of \(\tilde{w}_j\) is given by:

\begin{equation} \label{eqn:5}
\tilde{w}_j = \left\{1 + \frac{(1 - w_j) \sqrt{\kappa_j}}{w_j} \exp\left[-\frac{\beta_j^2}{2 \sigma^2} \left(\frac{\kappa_j - 1}{\kappa_j \zeta_j}\right)\right]\right\}^{-1}.
\end{equation}

In this study, \(\tilde{w}_j\), \(\kappa_j\), and \(\zeta_j\) are treated as constants, with values chosen to satisfy the constraints of the subsequent analyses.

\subsubsection{The Bayesian Lasso}
In the Bayesian lasso framework, each \(\tau_j\) is assigned an independent Exponential \(\left(\lambda^2 / 2\right)\) prior, where \(\lambda^2 / 2\) is the rate parameter of the exponential distribution. This setup implies that the marginal prior of \(\boldsymbol{\beta}\) (given \(\sigma^2\)) follows independent Laplace distributions for each component, effectively shrinking the regression coefficients. Consequently, the conditional posterior distribution of \(\boldsymbol{\tau} \mid (\boldsymbol{\beta}, \sigma^2, \boldsymbol{Y})\) assigns independent inverse Gaussian distributions to each \(1 / \tau_j\), which allows for efficient sampling. This full framework is extensively discussed by \cite{park2008bayesian} within the context of the 3BG algorithm.

\subsection{The Two-Block Gibbs Sampler for Bayesian Shrinkage Models}
\cite{Rajaratnam2019} introduce a lemma that forms the foundation for a two-block Gibbs sampler, which addresses the slow convergence properties of the three-block Gibbs sampler .

\noindent\textit{Lemma 1.} For the Bayesian model specified in equation \ref{eqn:2}, the posterior distribution of \(\sigma^2 \mid \boldsymbol{\tau}, \boldsymbol{Y}\) follows an inverse gamma distribution with shape parameter \((n-1)/2\) and scale parameter \(\tilde{\boldsymbol{Y}}^{\mathrm{T}}\left(\boldsymbol{I}_n-\boldsymbol{X} \boldsymbol{A}_{\boldsymbol{\tau}}^{-1} \boldsymbol{X}^{\mathrm{T}}\right) \tilde{\boldsymbol{Y}} / 2\).

This lemma enables the development of a 2BG sampler to generate samples from the joint posterior distribution of \((\boldsymbol{\beta}, \sigma^2)\), which retains the computational tractability of the original 3BG sampler. The key difference lies in the way \(\sigma^2\) is sampled: instead of drawing \(\sigma^2 \mid \boldsymbol{\beta}, \boldsymbol{\tau}, \boldsymbol{Y}\) as done in 3BG (refer to equation \ref{eqn:3}), the 2BG sampler uses the draw \(\sigma^2 \mid \boldsymbol{\tau}\), as derived in Lemma 1. The algorithm alternates between sampling \((\boldsymbol{\beta}, \sigma^2) \mid \boldsymbol{\tau}, \boldsymbol{Y}\) and \(\boldsymbol{\tau} \mid (\boldsymbol{\beta}, \sigma^2, \boldsymbol{Y})\). Specifically, \((\boldsymbol{\beta}, \sigma^2)\) is sampled by first drawing \(\sigma^2 \mid \boldsymbol{\tau}, \boldsymbol{Y}\), followed by drawing \(\boldsymbol{\beta} \mid \sigma^2, \boldsymbol{\tau}, \boldsymbol{Y}\).

The cyclical steps of the 2BG sampler can be summarized as follows:

\[
\begin{aligned}
& \boldsymbol{\tau} \mid \boldsymbol{\beta}, \sigma^2, \boldsymbol{Y} \sim \pi\left(\boldsymbol{\tau} \mid \boldsymbol{\beta}, \sigma^2, \boldsymbol{Y}\right), \\
& \left(\boldsymbol{\beta}, \sigma^2\right) \mid \boldsymbol{\tau}, \boldsymbol{Y} \sim \left\{
\begin{array}{l}
\sigma^2 \mid \boldsymbol{\tau}, \boldsymbol{Y} \sim \text{Inverse-Gamma}\left[\frac{n-1}{2}, \frac{\tilde{\boldsymbol{Y}}^{\mathrm{T}} \left(\boldsymbol{I}_n - \boldsymbol{X} \boldsymbol{A}_{\boldsymbol{\tau}}^{-1} \boldsymbol{X}^{\mathrm{T}}\right) \tilde{\boldsymbol{Y}}}{2}\right], \\
\boldsymbol{\beta} \mid \sigma^2, \boldsymbol{\tau}, \boldsymbol{Y} \sim \mathrm{N}_p\left(\boldsymbol{A}_{\boldsymbol{\tau}}^{-1} \boldsymbol{X}^{\mathrm{T}} \tilde{\boldsymbol{Y}}, \sigma^2 \boldsymbol{A}_{\boldsymbol{\tau}}^{-1}\right),
\end{array} \right.
\end{aligned}
\]

where \(\boldsymbol{A}_{\boldsymbol{\tau}} = \boldsymbol{X}^{\mathrm{T}} \boldsymbol{X} + \boldsymbol{D}_{\boldsymbol{\tau}}^{-1}\), and \(\boldsymbol{D}_{\boldsymbol{\tau}}\) is the diagonal matrix formed by the \(\tau_j\)'s.

\section{Simulation studies} \label{sec:simulation}
We assess the computational efficiency of the 2BG and 3BG samplers through a series of simulation studies. All simulations were conducted on a Windows 11 Pro PC with 16.0 GB RAM, 8 cores, and an Intel(R) Core(TM) i7-8650U @ 1.90GHz processor.

\subsection{Data Generation}

The data is simulated from the following linear model:
\[
    \boldsymbol{Y} = \boldsymbol{X} \boldsymbol{\beta}_* + \boldsymbol{\epsilon},
\]
where \(\boldsymbol{\epsilon} \in \mathbb{R}^{n \times 1}\) is a vector of standard normal random variables, and \(\boldsymbol{\beta}_* \in \mathbb{R}^{p \times 1}\) represents the true regression coefficients. We conduct simulations for two distinct models:

1. \textit{Spike-and-Slab Model}: Two sets of simulations are performed with \(n = \{50, 100\}\).

2. \textit{Bayesian Lasso Model}: A single set of simulations is performed with \(n = 75\).

For both models, the dimension \(p\) is chosen such that \(\frac{p}{n} \in \{0.5, 0.6, 0.7, 0.8, 0.9, 1, 2, 3, 4, 5\}\). This results in 10 datasets for each \((n, p)\) combination for the Spike-and-Slab model using the 2-block Gibbs sampler and 1-10 block simulations for the Bayesian Lasso model.

For each dataset, the design matrix \(\boldsymbol{X} \in \mathbb{R}^{n \times p}\) is generated by drawing rows independently from the \(p\)-dimensional standard multivariate normal distribution. The columns of \(\boldsymbol{X}\) are then standardized to have a mean of zero and a squared Euclidean norm of \(n\). For each \(p\), the first \(p/5\) elements of \(\boldsymbol{\beta}_*\) are nonzero, drawn independently from the \(t_2\) distribution.

\subsection{Simulation Setup}
We use the \textit{foreach} package in \textit{R} to run 6 parallel Markov chains, each with 15,000 iterations. The first 10\% of iterations are discarded as burn-in. All chains are initialized with \(\boldsymbol{\beta} = \boldsymbol{1}_p\) and \(\sigma_0^2 = 0\). The regularization parameter for the Bayesian Lasso model is set to \(\lambda = 1\). 

For the Spike-and-Slab model, the hyperparameters are consistently set as \(\tilde{w_j} = 1/2\), \(\kappa_j = 100\), and \(\zeta_j = 1/100\).

\subsection{Performance Metrics}
To evaluate computational efficiency, we estimate the average lag-1 autocorrelation \(\rho_1\) of the \(\sigma^2\) marginal after burn-in. This metric is chosen because \cite{rajaratnam2015mcmc} suggest that \(\sigma^2\) offers better insight into chain mixing than \(\boldsymbol{\beta}\). Additionally, we estimate the effective sample size per second, \(N_{\mathrm{eff}}/T\), where \(N_{\mathrm{eff}}\) is calculated using the R package \textit{coda} \citep{plummer2006coda} as:

\[
    N_{\mathrm{eff}} = \frac{N}{1 + 2 \sum_{k=1}^{\infty} \rho_k},
\]
with \(N\) denoting the total number of iterations and \(\rho_k\) the lag-\(k\) autocorrelation.

\subsection{Results for the \textit{Spike-and-slab} model} \label{ss_sim}
\begin{figure}[h]
         \centering
         \includegraphics[width=120mm]{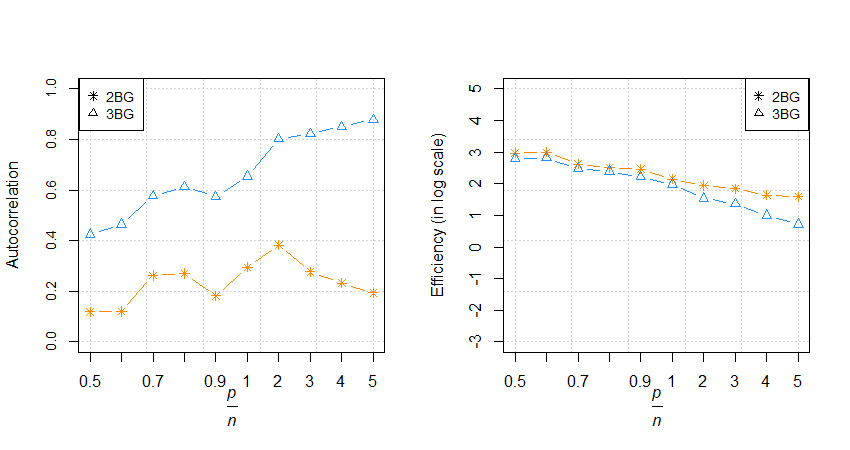}
         \caption{Empirical lag-one autocorrelation at $n=50$ (left) and the average effective sample size per second in base-10 log (right) of the $\sigma^2$ component of the four MCMC chains for the \textit{spike-and-slab} model.}
         \label{fig:ss50}
\end{figure}

The left panel of Figure \ref{fig:ss50} presents the empirical average lag-one autocorrelation of the \(\sigma^2\) component across the six MCMC chains for the \textit{spike-and-slab} model at \(n=50\). It is evident that the new 2BG sampler consistently exhibits smaller average autocorrelations compared to the 3BG sampler across all \((n, p)\) combinations. This suggests that the 2BG sampler has a better mixing rate, and thus, is computationally more efficient than the 3BG algorithm.

Moreover, as the dimensionality of the covariates \(p\) increases relative to the sample size \(n\), the performance gap between the 2BG and 3BG samplers widens, further emphasizing the efficiency of the 2BG sampler in high-dimensional settings.

The right panel of Figure \ref{fig:ss50} shows the average effective sample size per second, \(N_{\mathrm{eff}} / T\), on a base-10 log scale for each sampler. The results demonstrate that the 2BG sampler consistently produces more effective samples than the 3BG sampler. As \(p\) grows larger relative to \(n\), the computational advantage of the 2BG sampler becomes more pronounced. This suggests that the 2BG sampler incurs a smaller computational cost per chain, making it a more suitable algorithm for high-dimensional problems.

\begin{figure}[h]
         \centering
         \includegraphics[width=120mm]{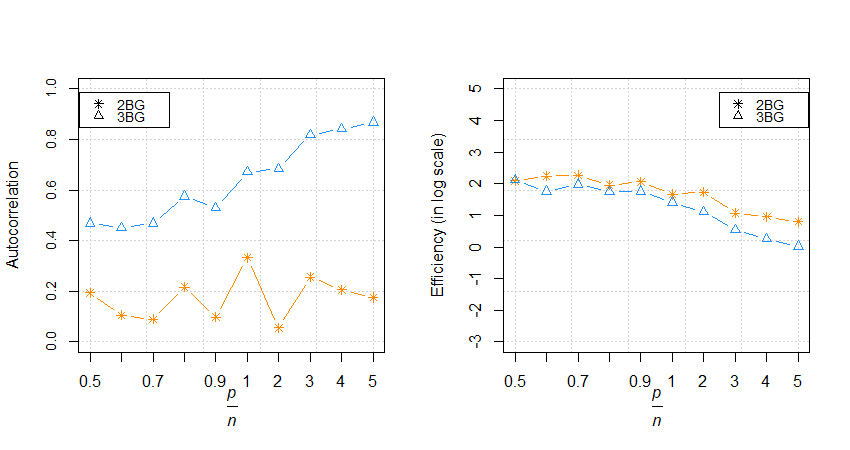}
         \caption{Empirical lag-one autocorrelation at $n=100$ (left) and the average effective sample size per second in base-10 log (right) of the $\sigma^2$ component of the four MCMC chains for the \textit{spike-and-slab} model}
         \label{fig:ss100}
\end{figure}

A similar trend is observed when scaling up the sample size to \(n = 100\), as depicted in Figure \ref{fig:ss100}. The 2BG sampler continues to outperform the 3BG in terms of both autocorrelation and effective sample size.

\subsection{Results for the Bayesian lasso model}
\begin{figure}[h]
         \centering
         \includegraphics[width=120mm]{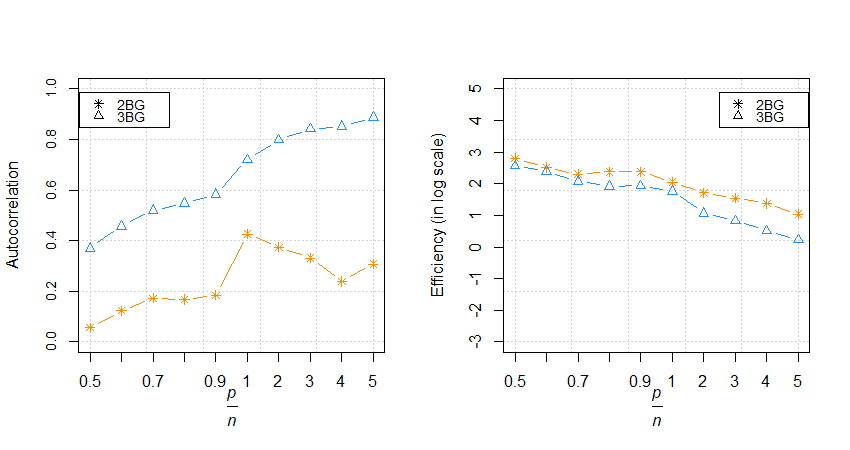}
         \caption{Empirical lag-one autocorrelation at $n=75$ (left) and the average effective sample size per second in base-10 log (right) of the $\sigma^2$ component of the four MCMC chains for the Bayesian lasso model}
         \label{fig:bl75}
\end{figure}

A similar analysis as in Section \ref{ss_sim} is performed under the Bayesian Lasso model. The left panel of Figure \ref{fig:bl75} presents the empirical average lag-one autocorrelation of the \(\sigma^2\) component across six MCMC chains at \(n = 75\). Consistent with the findings for the \textit{spike-and-slab} model, the 2BG sampler exhibits smaller average autocorrelations than the 3BG sampler across all \((n, p)\) combinations. This trend holds under both the "$n$-small, $p$-large" and "$n$-large, $p$-small" regimes.

Interestingly, the closest average lag-one autocorrelation between the two algorithms occurs when \(n = p = 100\), suggesting comparable performance in this specific scenario. However, as \(p\) increases relative to \(n\), the 2BG sampler consistently outperforms the 3BG sampler in terms of mixing efficiency.

The right panel of Figure \ref{fig:bl75} shows the average effective sample size per second, \(N_{\mathrm{eff}} / T\), also on a base-10 log scale. Unsurprisingly, the 2BG sampler generates significantly more effective samples than the 3BG counterpart, further demonstrating its computational superiority. This indicates that the 2BG sampler achieves a higher degree of efficiency, especially in high-dimensional settings, making it a more suitable choice for Bayesian Lasso models.

\section{Application to Real Data} \label{sec:application}
We now apply both the Bayesian lasso and the \textit{spike-and-slab} models to a real dataset. We use the well-known \href{https://discover.nci.nih.gov/cellminer/}{NCI-60 cancer cell panel} from the National cancer Institute. This dataset comprises protein expressions for a specific protein selected as the response variable, and the gene expressions of the 100 genes that have the highest (robustly estimated) correlations with the response variable which are screened as candidate predictors. This pre-processing is done using the R \textit{robustHD} package \citep{alfons2021robusthd}. Hence, we have $n=59$ samples and $p=100$ covariates. Each covariate was further standardized to have mean zero and squared Euclidean norm $n$. We set the hyperparameters of the priors $\tilde{w_j} = 1/2, \kappa = 100$ and $\zeta = 1/200$ with $\lambda = 0.5$. For both shrinkage models, we run 18,000 chains and set the first 10\% aside as burn-in.

\begin{figure}[h]
         \centering
         \includegraphics[width=100mm]{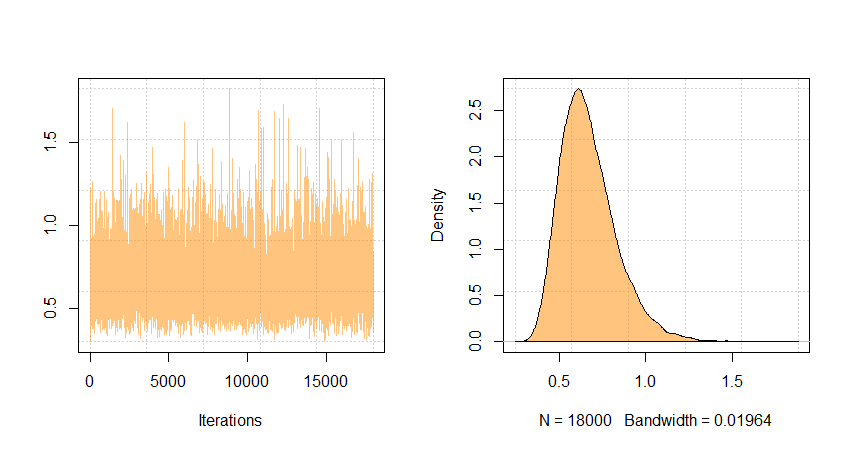}
         \caption{Trace-plot and density plot of the 2BG \textit{spike-and-slab} model for the proteins data}
         \label{fig:ssreal}
\end{figure}

\begin{figure}[h]
         \centering
         \includegraphics[width=100mm]{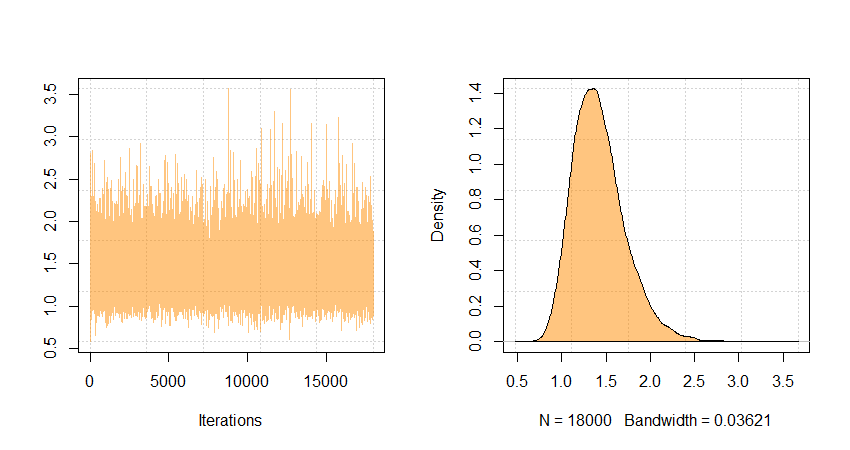}
         \caption{Trace-plot and density plot of the 2BG Bayesian lasso model for the proteins data}
         \label{fig:blreal}
\end{figure}

As illustrated in Figures \ref{fig:ssreal} and \ref{fig:blreal}, the chains for both the \textit{spike-and-slab} and Bayesian lasso models have attained stationarity, indicating that they provide sufficient samples for posterior estimation. Further, from Table \ref{table:gene}, we observe that the lag-one autocorrelation for the 2BG model for both the \textit{spike-and-slab} and Bayesian lasso models (0.387 and 0.161 respectively) are much smaller than the 3BG counterparts, indicating better mixing. 

Moreover, the 2BG sampler generates about twice as many effective samples as the 3BG sampler in the \textit{spike-and-slab} model, and approximately five times as many for the Bayesian Lasso model. This demonstrates that the 2BG sampler is not only computationally more efficient but also highly effective, particularly in high-dimensional scenarios.

\begin{table}[h!]
\centering
\begin{tabular}{ p{3.5cm} p{1.5cm} p{1.5cm}| p{1.5cm} p{1.5cm} |p{1.5cm} p{1.5cm} }
  \hline
  \multicolumn{3}{c}{ } & 
  \multicolumn{2}{l}{\underline{Autocorrelation}\rlap{\underline{~}}} &
  \multicolumn{2}{c}{\underline{$N_{\mathrm{eff}}$}\rlap{\underline{}}}  \\
  Dataset & $n$ & $p$ &  2GB & 3GB & 2BG & 3BG \\
  \hline
  Gene (\textit{spike-and-slab}) & 59 & 100 &   0.387 & 0.774  & 3,263 & 1,639 \\
Gene (\textit{Bayesian lasso}) & 59 & 100 & 0.161 & 0.700 & 10,921 & 2,856 \\
  \hline
\end{tabular}
\caption{Lag-one autocorrelation and effective sample size per second for the $\sigma^2$-component of 2BG and 3BG of the \textit{spike-and-slab} and Bayesian Lasso models as applied to the protein gene dataset}
\label{table:gene}
\end{table}

\section{Discussion} \label{sec:discussion}
In this study, we have used both simulation studies and real data analysis to demonstrate the computational prowess of the two-block Gibbs (2BG) sampler algorithm, specifically within the context of the \textit{spike-and-slab} and Bayesian Lasso models. Our results consistently show that the 2BG is faster and more efficient relative to the three-block Gibbs (3BG) sampler. This efficiency is particularly evident in its superior mixing rates, as evidenced by lower lag-one autocorrelations, and its ability to produce substantially higher effective sample sizes per unit of time. These computational advantages make the 2BG sampler a highly attractive choice for practitioners working in high-dimensional settings, where computational complexity can often be a bottleneck.

Furthermore, this study extends the application to real-world data from the NCI-60 cancer cell line panel, using gene expression data to predict protein expression levels. By incorporating feature selection, the analysis not only identifies the most influential genes but also provides insights into the genetic mechanisms driving protein expression in cancer cells. This contribution deepens the understanding of shrinkage methods in high-dimensional settings while offering novel applications in cancer genomics. The results of our real data application show that the 2BG sampler can handle real-world, high-dimensional data efficiently, providing both robust estimates and computational scalability.

\clearpage

\bibliographystyle{natbib}

\bibliography{template}

\begin{thebibliography}{}

\bibitem[Alfons(2021)Alfons]{alfons2021robusthd}
Alfons, A. (2021).
\newblock robusthd: An r package for robust regression with high-dimensional
  data.
\newblock {\em Journal of Open Source Software\/}, {\bf 6}(67), 3786.

\bibitem[Carvalho {\em et~al.}(2010)Carvalho, Polson, and
  Scott]{carvalho2010horseshoe}
Carvalho, C.~M.  {\em et~al.} (2010).
\newblock The horseshoe estimator for sparse signals.
\newblock {\em Biometrika\/}, {\bf 97}(2), 465--480.

\bibitem[Jin and Tan(2021)Jin and Tan]{jin2021fast}
Jin, R. and Tan, A. (2021).
\newblock Fast markov chain monte carlo for high-dimensional bayesian
  regression models with shrinkage priors.
\newblock {\em Journal of Computational and Graphical Statistics\/}, {\bf
  30}(3), 632--646.

\bibitem[Johnstone and Silverman(2004)Johnstone and
  Silverman]{johnstone2004needles}
Johnstone, I.~M. and Silverman, B.~W. (2004).
\newblock Needles and straw in haystacks: Empirical bayes estimates of possibly
  sparse sequences.

\bibitem[Khare and Hobert(2013)Khare and Hobert]{khare2013geometric}
Khare, K. and Hobert, J.~P. (2013).
\newblock Geometric ergodicity of the bayesian lasso.

\bibitem[Neville {\em et~al.}(2014)Neville, Ormerod, and Wand]{neville2014mean}
Neville, S.~E.  {\em et~al.} (2014).
\newblock Mean field variational bayes for continuous sparse signal shrinkage:
  pitfalls and remedies.

\bibitem[Park and Casella(2008)Park and Casella]{park2008bayesian}
Park, T. and Casella, G. (2008).
\newblock The bayesian lasso.
\newblock {\em Journal of the American Statistical Association\/}, {\bf
  103}(482), 681--686.

\bibitem[Plummer {\em et~al.}(2006)Plummer, Best, Cowles, Vines, {\em
  et~al.}]{plummer2006coda}
Plummer, M.  {\em et~al.} (2006).
\newblock Coda: convergence diagnosis and output analysis for mcmc.
\newblock {\em R news\/}, {\bf 6}(1), 7--11.

\bibitem[Rajaratnam and Sparks(2015)Rajaratnam and Sparks]{rajaratnam2015mcmc}
Rajaratnam, B. and Sparks, D. (2015).
\newblock Mcmc-based inference in the era of big data: A fundamental analysis
  of the convergence complexity of high-dimensional chains.
\newblock {\em arXiv preprint arXiv:1508.00947\/}.

\bibitem[Rajaratnam {\em et~al.}(2019)Rajaratnam, Sparks, Khare, and
  Zhang]{Rajaratnam2019}
Rajaratnam, B.  {\em et~al.} (2019).
\newblock Uncertainty quantification for modern high-dimensional regression via
  scalable bayesian methods.
\newblock {\em Journal of Computational and Graphical Statistics\/}, {\bf
  28}(1), 174--184.

\bibitem[Tibshirani(1996)Tibshirani]{tibshirani1996regression}
Tibshirani, R. (1996).
\newblock Regression shrinkage and selection via the lasso.
\newblock {\em Journal of the Royal Statistical Society: Series B
  (Methodological)\/}, {\bf 58}(1), 267--288.

\end{thebibliography}

\end{document}